\documentclass[prl,twocolumn,superscriptaddress]{revtex4}%
\usepackage{amsfonts}
\usepackage{amsmath}
\usepackage{amssymb}
\usepackage{hyperref}
\usepackage{color}
\usepackage{graphicx}%
\providecommand{\U}[1]{\protect\rule{.1in}{.1in}}

\newenvironment{proof}[1][Proof]{\noindent\textbf{#1.} }{\ \rule{0.5em}{0.5em}}

\let\originalleft\left
\let\originalright\right
\renewcommand{\left}{\mathopen{}\mathclose\bgroup\originalleft}
\renewcommand{\right}{\aftergroup\egroup\originalright}

\begin{document}
\preprint{ }
\title[ ]{Fundamental rate-loss tradeoff for optical quantum key distribution}
\author{Masahiro Takeoka}
\affiliation{National Institute of Information and Communications Technology, Koganei, Tokyo 184-8795, Japan}
\affiliation{Quantum Information Processing Group, Raytheon BBN Technologies, Cambridge, MA 02138, USA}
\author{Saikat Guha}
\affiliation{Quantum Information Processing Group, Raytheon BBN Technologies, Cambridge, MA 02138, USA}
\author{Mark M.~Wilde}
\affiliation{Hearne Institute for Theoretical Physics, Department of Physics and Astronomy,
Center for Computation and Technology, Louisiana State University, Baton
Rouge, Louisiana 70803, USA}

\begin{abstract}
Since 1984, 
various optical quantum key distribution (QKD) 
protocols have been proposed and examined. In all of them, 
the rate of secret key generation decays exponentially with distance. 
A natural and fundamental question is then whether there are yet-to-be 
discovered optical QKD protocols (without quantum repeaters) 
that could circumvent this rate-distance tradeoff. 
This paper provides a major step towards answering this question. 
We show that the secret-key-agreement capacity of a lossy and noisy 
optical channel assisted by unlimited two-way public classical communication 
is limited by an upper bound that is solely a function of the channel loss, 
regardless of how much optical power the protocol may use. 
Our result has major implications for understanding the secret-key-agreement 
capacity of optical channels---a long-standing open problem in optical quantum 
information theory---and strongly suggests a real need for quantum repeaters 
to perform QKD at high rates over long distances.
\end{abstract}
\volumeyear{2013}
\volumenumber{ }
\issuenumber{ }
\eid{ }
\startpage{1}
\endpage{10}
\maketitle

The goal of quantum key distribution (QKD) is to generate a shared secret key between two distant parties Alice and Bob, such that the key is perfectly secret from any eavesdropper, Eve. Since the invention of the BB84 protocol \cite{BB84}, the theory and practice of QKD has come a long way. Various different QKD protocols have been proposed in the last three decades~\cite{SBCDLP09}, some of which are now turning from science into practical technologies~\cite{SECOQC,TOKYO_QKD,Elliott02}. Security of QKD has now been proven for many protocols and under practical limitations such as a finite key length~\cite{TLGR12,FFBLSTR12}. 

It is well recognized that the key rates of all known QKD protocols (such as BB84~\cite{BB84}, E91~\cite{E91}, CV-GG02~\cite{GG02}) decay exponentially with distance. To obtain a large key rate across a long distance link, the link can be divided into many low-loss segments separated by trusted (physically-secured) relay nodes. Interestingly however, quantum mechanics permits building QKD protocols using devices called quantum repeaters, which if supplied at the relay nodes, would make it unnecessary to physically secure them, thus enabling long-distance high-rate QKD. 

Quantum repeater technology, in particular ones built using quantum memories, have been a subject of intense investigation in recent years~\cite{SST_2011,LST_2009}, but an operational demonstration of a quantum repeater has proven extremely challenging and has yet to be conducted 
(note that some repeater protocols without quantum memory have been proposed recently \cite{MSDHN12,ATL13}, but their implementation is still challenging).

The natural question that thus arises, is whether there are yet-to-be-discovered optical QKD protocols that could circumvent the exponential rate-distance tradeoff of BB84 and transmit at high rates over long distances without the need for quantum repeaters. 
In this paper, we establish that it is not possible to do so.
We employ information-theoretic techniques in order to establish our main result.

The secret key agreement capacity of a quantum channel
is the highest rate (bits per channel use) at which a shared key 
can be reliably and securely generated using the channel many times 
in conjunction with unlimited 
two-way classical communication over an authenticated
public channel. This paper establishes that,
regardless of the QKD protocol used, a fundamental limit on
the secret key agreement capacity of a lossy optical channel is given by an upper bound that is solely a function of the channel loss, i.e., independent of the transmit power.
We show that the bound is nearly optimal at high loss, the regime relevant for practical QKD. We also compare our upper bound and the best-known achievable rate with the ideal BB84 and CV-GG02 protocols. We find that even though there is room for improvement over these protocols, there is essentially no gap in the rate-loss scaling. 
We thereby place on a firm foundation 
the need for quantum repeaters for high rate QKD over long distances 
with no trusted relays.
We note that the upper bound proved here is a so-called `weak converse' 
upper bound, meaning that if the communication rate of any secret key agreement
protocol exceeds this bound, then our theorem implies that its reliability 
and security can never be perfect, even in the asymptotic limit of 
many channel uses (we will also discuss the issue of a finite number of 
channel uses below and the Methods section).

\begin{figure}[!t]
\centering
\includegraphics[width=3.0in]{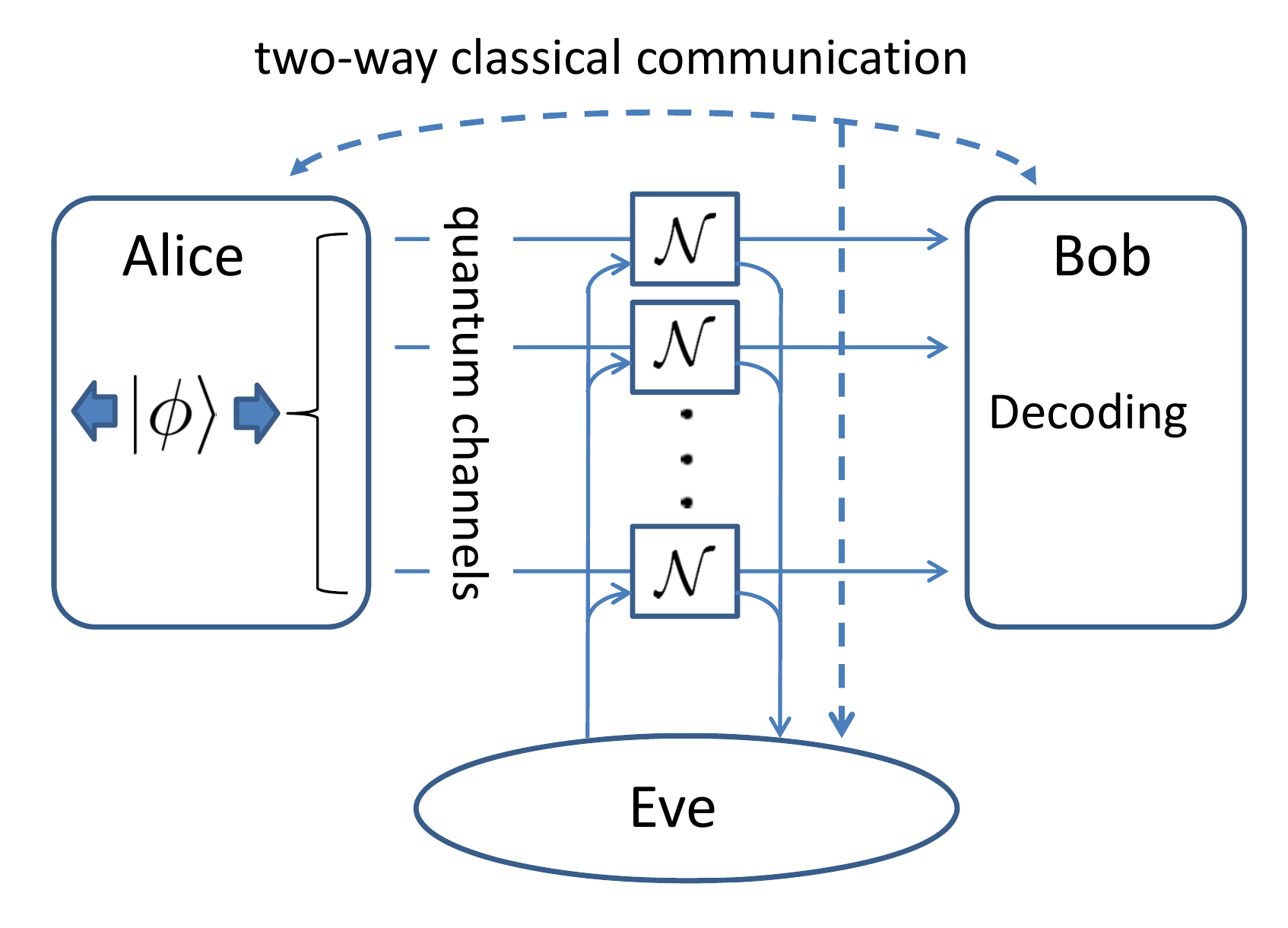}
\caption{{\bf A generic point-to-point QKD protocol}. Alice transmits quantum state via $n$ forward uses of a quantum channel to Bob. They are also allowed to use unlimited forward and backward public classical communication over an authenticated classical channel.}
\label{schematic}
\end{figure}

A generic point-to-point QKD protocol 
is illustrated in Fig.~\ref{schematic}. In the protocol,
the sender Alice transmits over $n$ independent uses
of the quantum channel ${\cal N}$. The legitimate receiver Bob
obtains the outputs of the channel uses. Note that Alice
could send product or entangled states. They are also allowed unlimited two-way public classical communication over an authenticated channel, in order to generate shared secret key. The adversary Eve is assumed to be `all-powerful': one who has access to the full environment of the Alice-to-Bob quantum channel, the `maximal' quantum system to which she can have access. Optically, this translates to Eve being able to collect every single photon that does not enter Bob's receiver. Eve may also actively attack, for instance by injecting a quantum state into the channel. She has access to all the public classical communication between Alice and Bob. Finally, Eve is assumed to be able to store the quantum states she obtains over all $n$ channel uses without any loss or degradation, and she can make any collective quantum measurement on those systems, in an attempt to learn the secret key. In the theory of QKD, one attributes all channel
impairments (such as loss, noise, turbulence, detector imperfections) collectively measured by Alice and Bob during a channel-estimation step of the protocol, to adversarial actions of the worst case Eve with which the measured channel is `consistent' (even though in reality all those impairments may have been caused by non-adversarial natural phenomena). Alice and Bob then run a key generation protocol aiming to generate secret key at a rate close to the secret key agreement capacity of that channel. It is hard in general to calculate this capacity precisely, and even more so to come up with protocols that can attain key rates close to that capacity. 

In this paper, we show a strong limitation on the secret key agreement 
capacity of a general memoryless quantum channel. 
More precisely, we provide a simple upper bound on the two-way assisted private capacity $P_2(\mathcal{N})$ of a quantum channel~$\mathcal{N}$. Note that the capacity for secret-key agreement using QKD as discussed above is equal to the capacity for private communication with unlimited public discussion, due to the one-time pad protocol. We first define a new quantity, the squashed entanglement of a quantum channel, and show that it is a simple upper bound on $P_2(\mathcal{N})$ for any quantum channel $\mathcal{N}$. 
We then apply it to the pure-loss optical channel $\mathcal{N}_\eta$ with transmittance $\eta \in [0,1]$. Channel loss is an important and measurable impairment both for free-space and fiber optical links, and is directly tied to the communication distance (for example, a low-loss fiber may have a loss of $0.2$ dBkm$^{-1}$). Since any excess noise in the channel or detectors can only reduce the extractable key rate, our upper bound on $P_2(\mathcal{N}_\eta)$ imposes a fundamental upper limit on the secret key rate of any point-to-point QKD protocol over a lossy optical channel not assisted by any quantum repeater. We will establish the following upper bound: 
\begin{equation}
\label{eq:good-upper-bound}
P_2(\mathcal{N}_\eta) \le \log_2 ((1+\eta)/(1-\eta))\; {\text{key bits per mode}}.
\end{equation}
In the practical regime of high loss ($\eta \ll 1$), we will argue that this upper bound is only a factor of two higher than the best known lower bound, $P_2(\mathcal{N}_\eta) \ge \log\left(1/(1-\eta)\right)$~\cite{GPLS09}. We will also extend our upper bound to the more general scenario of QKD using a two-way quantum channel with unlimited public discussion, which was discussed recently in Ref.~\cite{PMLB08}. Finally, we will compare our upper bound and the best-known lower bound to the rate achievable by the BB84 and the CV-GG02 protocols under best-case operating conditions as well as compare to the performance of these protocols under realistic operating conditions. 

\section*{Results} 
{\bf Squashed entanglement.}
Before discussing our main results, we briefly review the squashed 
entanglement, which plays an important role in our work. 
The secret-key agreement capacity assisted by public communication was defined for a classical channel $p_{Y,Z|X}$ ($X=$ Alice, $Y=$ Bob, $Z=$ Eve), independently by Maurer~\cite{M93}, and Ahlswede and Csisz{\'{a}}r~\cite{AC93}, who proved lower and upper bounds on the capacity. Maurer and Wolf later introduced the intrinsic information $I(X;Y{\downarrow}Z) \equiv \min\left\{I(X;Y|Z^\prime): P_{X,Y,Z,Z^\prime} = P_{X,Y,Z}P_{Z^\prime |Z}\right\}$, and proved that this quantity optimized over all channel input distributions is a sharp upper bound on the secret key agreement capacity of 
$p_{Y,Z|X}$ \cite{MW99}. Leveraging strong parallels discovered between secrecy and quantum coherence~\cite{SW98,LC99,SP00}, Christandl and Winter extended the intrinsic information quantity to the
realm of quantum information theory. They defined the squashed entanglement $E_{\text{sq}}\left(A;B\right) _{\rho}$\ of a bipartite quantum state $\rho_{AB}$, and proved it to be an upper bound on the rate at which two parties can distill maximally entangled (Bell) states $\left(\left\vert 0\right\rangle \left\vert 0\right\rangle +\left\vert 1\right\rangle\left\vert 1\right\rangle \right)  /\sqrt{2}$ from many copies of $\rho_{AB}$ using local operations and classical communication (LOCC)~\cite{CW04}. Using a similar technique, the squashed entanglement was proved to upper bound the distillable secret key rate~\cite{CEHHOR07,C06}. The squashed entanglement of a bipartite state $\rho_{AB}$ is defined as
\begin{equation}
E_{\text{sq}}\left(  A;B\right)  _{\rho}\equiv\tfrac{1}{2}\inf_{\mathcal{S}%
_{E\rightarrow E^{\prime}}}I\left(  A;B|E^{\prime}\right)  ,
\label{eq:squashed-ent-state}%
\end{equation}
where $I(A;B|E^{\prime}) \equiv H(AE') + H(BE') - H(E') - H(ABE')$ is the conditional quantum mutual information and the infimum is taken over all noisy `squashing channels' $\mathcal{S}_{E\rightarrow E^{\prime}}$
taking the $E$ system of a purification $\left\vert \phi^{\rho}\right\rangle_{ABE}$\ of $\rho_{AB}$ to a system $E^{\prime}$ of arbitrary dimension.
In related work, Tucci has defined a functional bearing some similarities 
to squashed entanglement \cite{T99,T02}. We can interpret $E_{\text{sq}}\left(  A;B\right)  _{\rho}$ as quantifying the minimum remnant quantum correlations between $A$ and $B$ after an adversary possessing the purifying system $E$ performs a quantum operation on it with the intent of `squashing down' the correlations that $A$ and $B$ share. It should also be noted that among the many entanglement measures, squashed entanglement is the only one known to satisfy all eight desirable properties that have arisen in the axiomatization of entanglement theory \cite{CW04,KW04,AF04,BCY11}.

{\bf Squashed entanglement of a quantum channel.}
The upper bound from \cite{CEHHOR07} on the distillable key rate applies 
to the scenario in which Alice and Bob share many copies
of some bipartite state $\rho_{AB}$. 
In order to upper bound the key agreement capacity of a channel, 
we define the squashed entanglement of a quantum channel $\mathcal{N}_{A^{\prime}\rightarrow B}$ as the maximum squashed entanglement that can be registered between a sender and receiver with access to the input $A^{\prime}$\ and output $B$\ of this channel,
respectively:%
\begin{equation}
E_{\text{sq}}\left(  \mathcal{N}\right)  \equiv\max_{\left\vert \phi
\right\rangle _{AA^{\prime}}}E_{\text{sq}}\left(  A;B\right)  _{\rho},
\label{eq:squashed-ent-channel}%
\end{equation}
where $\rho \equiv \rho_{AB}\equiv\mathcal{N}_{A^{\prime}\rightarrow B}(\left\vert
\phi\right\rangle \left\langle \phi\right\vert _{AA^{\prime}})$.
Note that, in the above formula, we can take a maximum rather than a supremum 
if the input space is finite-dimensional because in this case, the input 
space is compact and the squashed entanglement measure is continuous 
\cite{AF04}. Also, we can restrict the optimization 
to be taken over pure bipartite states, due to the convexity of squashed 
entanglement \cite{CW04}.

We now prove that $E_{\text{sq}}\left(  \mathcal{N}\right)$
plays an operational role analogous to intrinsic information, i.e., it upper bounds the secret-key agreement capacity $P_{2}\left(  \mathcal{N}\right)  $.

{\it Theorem 1}: $E_{\operatorname{sq}}\left(  \mathcal{N}\right)  $ is
an upper bound on $P_{2}\left(  \mathcal{N}\right)  $, the private 
capacity of $\mathcal{N}$\ assisted by unlimited forward and 
backward classical communication:%
\begin{equation}
\label{eq:squashed_ent_upper_bound}
P_{2}\left(  \mathcal{N}\right)  \leq E_{\operatorname{sq}}\left(
\mathcal{N}\right)  .
\end{equation}

{\it Proof}: 
First recall that the squashed entanglement is a secrecy monotone, 
that is, it does not increase under local operations and public 
classical communication (LOPC) in the sense that 
$E_{\text{sq}}\left(A;B\right)_{\rho}\geq 
E_{\text{sq}}\left(A;B\right)_{\sigma}$ if Alice and Bob can obtain 
the state $\sigma_{AB}$ from $\rho_{AB}$ by LOPC \cite{C06,CEHHOR07}.
The method for doing so was to exploit the fact that LOPC distillation of
secret key is equivalent to LOCC distillation of private states 
\cite{HHHO05,HHHO09}. 
A private state has the following form \cite{HHHO05,HHHO09}: 
\[
\gamma_{ABA'B'} = U_{ABA'B'} \left( |\Phi\rangle\langle\Phi|_{AB} 
\otimes \rho_{A'B'} \right) U_{ABA'B'}^\dagger ,
\]
where $U_{ABA'B'}=\sum_{i,j} |i \rangle\langle i|_A \otimes 
|j \rangle\langle j|_B \otimes U^{ij}_{A'B'}$ is a global unitary operation, 
$|\Phi\rangle_{AB} = \sum_i |i\rangle_A |i\rangle_B / \sqrt{d}$ 
is a maximally entangled state of Schmidt rank $d$, 
and $\{|i\rangle_{A}\}$ and $\{|i\rangle_{B}\}$ are complete orthonormal
bases for quantum systems $A$ and $B$, respectively. 
Furthermore, the squashed entanglement is normalized,
in the sense that $E_{\text{sq}}\left(A;B\right)_{\gamma} \ge \log d$ 
(see Proposition 4.19 of \cite{C06}). 
Finally, the squashed entanglement satisfies the following continuity 
inequality \cite{AF04,C06}:%
\begin{multline}
\label{eq:continuity}
\text{if \ \ \ }\left\Vert \rho_{AB}-\sigma_{AB}\right\Vert _{1}%
\leq\varepsilon,\text{ \ \ \ then}\\
\left\vert E_{\text{sq}}\left(  A;B\right)  _{\rho}-E_{\text{sq}}\left(
A;B\right)  _{\sigma}\right\vert \leq16\sqrt{\varepsilon}\log d'+4h_{2}\left(
2\sqrt{\varepsilon}\right)  ,
\end{multline}
where $d'=\min\left\{  \left\vert A\right\vert ,\left\vert B\right\vert
\right\}  $ and $h_{2}\left(  x\right)  $ is the binary entropy function with
the property that $\lim_{x\rightarrow0}h_{2}\left(  x\right)  =0$. 

The most general $(n,R,\varepsilon)$ protocol in this setting 
is described as follows, where $n$ is the number of channel uses, 
$R$ is the key generation rate (measured in secret key bits per channel use), 
and $\varepsilon$ is a parameter quantifying the security (see below for 
their formal definitions). 
The protocol begins 
with Alice preparing a state $\rho_{AA_{1}\cdots A_{n}}^{\left(  1\right)  }$
on $n+1$ systems. She then transmits the system $A_{1}$ through one use of the
channel $\mathcal{N}$, and considering its isometric extension $U_{A_{1}%
\rightarrow B_{1}E_{1}}^{\mathcal{N}}$, we write the output state as
$\sigma_{AB_{1}E_{1}A_{2}\cdots A_{n}}^{\left(  1\right)  }$. Let $R^{(1)}$ be
a system that purifies this state. There is then a round of an arbitrary
amount of LOPC\ between Alice and Bob, resulting in a state $\rho_{AB_{1}%
E_{1}A_{2}\cdots A_{n}}^{\left(  2\right)  }$. This procedure continues, with
Alice transmitting system $A_{2}$ through the channel, leading to a state
$\sigma_{AB_{1}E_{1}B_{2}E_{2}A_{3}\cdots A_{n}}^{\left(  2\right)  }$, etc.
After the $n$th channel use, the state is $\sigma_{AB_{1}E_{1}B_{2}E_{2}\cdots
B_{n}E_{n}}^{\left(  n\right)  }$ (note that the dimension of the system $A$
might change throughout the protocol). Let $R^{(n)}$ be a reference system
that purifies this state. 
There is a final round of LOPC, producing a state $\omega
_{ABE_{1}\cdots E_{n}}$, whose reduction $\omega_{AB}$ satisfies 
\[
\left\Vert \omega_{AB}-\gamma_{AB}
\right\Vert _{1}\leq\varepsilon,
\]
where $\gamma_{AB}$ is a private state of $nR$ bits. Note that we are implicitly 
including the systems $A'$ and $B'$ in $A$ and $B$, respectively.

We can now proceed by bounding the secret key  
generation rate of any such protocol as follows:%
\begin{align*}
nR  &  \leq E_{\text{sq}}\left(  A;B\right)  _{\gamma}\\
&  \leq E_{\text{sq}}\left(  A;B\right)  _{\omega}+nf\left(  \varepsilon
\right)  .
\end{align*}
The first inequality follows from the normalization of the squashed entanglement on
private states (as mentioned above). The second inequality follows from the
continuity of squashed entanglement, with an appropriate choice of $f\left(
\varepsilon\right)$ so that $\lim_{\varepsilon\rightarrow0}f\left(
\varepsilon\right)=0$ (see the Methods section for more details). 
To continue, we introduce the following new subadditivity inequality for 
the squashed entanglement:

{\it Lemma 2}: 
For any five-party pure state $\psi_{AB_1E_1B_2E_2}$, 
\begin{equation}
E_{\operatorname{sq}}\left(  A;B_{1}B_{2}\right)  _{\psi} \leq
E_{\operatorname{sq}}\left(  AB_{2}E_{2};B_{1}\right)  _{\psi} 
+ E_{\operatorname{sq}}\left(  AB_{1}E_{1};B_{2}\right)  _{\psi}.\nonumber
\end{equation}

{\it Proof}: See the Supplementary Note 1 for a proof. 

With this new inequality in hand, we can establish the following chain of inequalities: 
\begin{align*}
E_{\text{sq}}\left(  A;B\right)  _{\omega}  &  \leq E_{\text{sq}}\left(
A;B_{1}\cdots B_{n}\right)  _{\sigma^{\left(  n\right)  }}\\
&  \leq E_{\text{sq}}(AB_{1}E_{1}\cdots B_{n-1}E_{n-1}R^{(n)};B_{n}%
)_{\sigma^{\left(  n\right)  }}\\
&  \ \ \ \ \ +E_{\text{sq}}\left(  AB_{n}E_{n};B_{1}\cdots B_{n-1}\right)
_{\sigma^{\left(  n\right)  }}\\
&  \leq E_{\text{sq}}\left(  \mathcal{N}\right)  +E_{\text{sq}}\left(
AB_{n}E_{n};B_{1}\cdots B_{n-1}\right)  _{\sigma^{\left(  n\right)  }}\\
&  =E_{\text{sq}}\left(  \mathcal{N}\right)  +E_{\text{sq}}\left(
AA_{n};B_{1}\cdots B_{n-1}\right)  _{\rho^{\left(  n\right)  }}\\
&  \leq nE_{\text{sq}}\left(  \mathcal{N}\right)  .
\end{align*}
The first inequality follows from monotonicity of the squashed entanglement
under LOCC. The second inequality is an application of the subadditivity
inequality in Lemma 2. The third inequality follows because
$E_{\text{sq}}(AB_{1}E_{1}\cdots B_{n-1}E_{n-1}R^{(n)};B_{n})_{\sigma^{\left(
n\right)  }}\leq E_{\text{sq}}\left(  \mathcal{N}\right)  $ (there is a
particular input to the $n$th channel, while the systems $AB_{1}E_{1}\cdots
B_{n-1}E_{n-1}R^{(n)}$ purify the system being input to the channel). The sole
equality follows because the squashed entanglement is invariant under local
isometries (the isometry here being the isometric extension of the channel).
The last inequality follows by induction, i.e., repeating this procedure by
using secrecy monotonicity and subadditivity, \textquotedblleft peeling
off\textquotedblright\ one term at a time. Putting everything together, we
arrive at
\[
nR\leq nE_{\text{sq}}\left(  \mathcal{N}\right)  +nf\left(  \varepsilon
\right)  ,
\]
which we can divide by $n$ and take the limit as $\varepsilon\rightarrow0$ to
recover the result that $P_{2}\left(\mathcal{N}\right)\leq E_{\text{sq}}
\left(  \mathcal{N}\right)$.
This completes the proof of Theorem 1. 

It should be stressed 
that the right hand of \eqref{eq:squashed_ent_upper_bound} is 
a `single-letter' expression, meaning that the expression is a function
of a single channel use. This is in spite of the fact that the quantity  
serves as an upper bound on the secret key agreement capacity, which
involves using the channel many independent times, 
entangled input states, and/or measurements over many channel outputs. 
Lemma 2 is critical for establishing 
the `single-letterization.'
The simple expression in \eqref{eq:squashed_ent_upper_bound}
allows us to apply the bound to various channels, 
including the optical channel as shown below. 

Also, as mentioned in the introduction, Theorem 1 states that 
$E_{\rm sq}(\mathcal{N})$ is a weak converse upper bound which 
bounds the key rate in the asymptotic limit of many channel uses. 
However, our bound is also valid for any finite number of channel uses,
in the sense that the key rate 
is upper bounded in terms of $E_{\rm sq}(\mathcal{N})$ and 
the reliability and security of the protocol. 
It might be possible to improve upon our upper bound, by establishing
a so-called strong converse theorem (see, e.g., \cite{ON99,W99}) 
or a refined second-order analysis, along the lines of \cite{TH12}, 
which is left as an important open question. 
We point the reader  to the Methods section for a quantitative discussion 
and an example scenario involving a pure-loss optical channel. 

A variation of this setting is one in which there is a forward quantum channel $\mathcal{N}$\ from Alice to Bob and a backward quantum channel $\mathcal{M}$\ from Bob to Alice. The most general protocol for generating a shared secret will have Alice and Bob each prepare a state on $n$ systems, Alice
sending one system through the forward channel, them conducting a round of LOPC, Bob sending one of his systems through the backward channel, them conducting a
round of LOPC, etc. Using essentially the same proof technique as above, it follows that $E_{\text{sq}}\left(  \mathcal{N}\right)  +E_{\text{sq}}\left(\mathcal{M}\right)  $ serves as an upper bound on the total rate at which Alice and Bob can generate a shared secret key using these two channels many independent times. It is also worth noting that $E_{\text{sq}}(\mathcal{N})$ is an upper bound on the two-way assisted quantum capacity $Q_2(\mathcal{N})$ (defined in \cite{BDS97}) because $P_2(\mathcal{N}) \ge Q_2(\mathcal{N})$ holds in general.

\begin{figure}
\centering
\includegraphics[width=0.6\columnwidth]{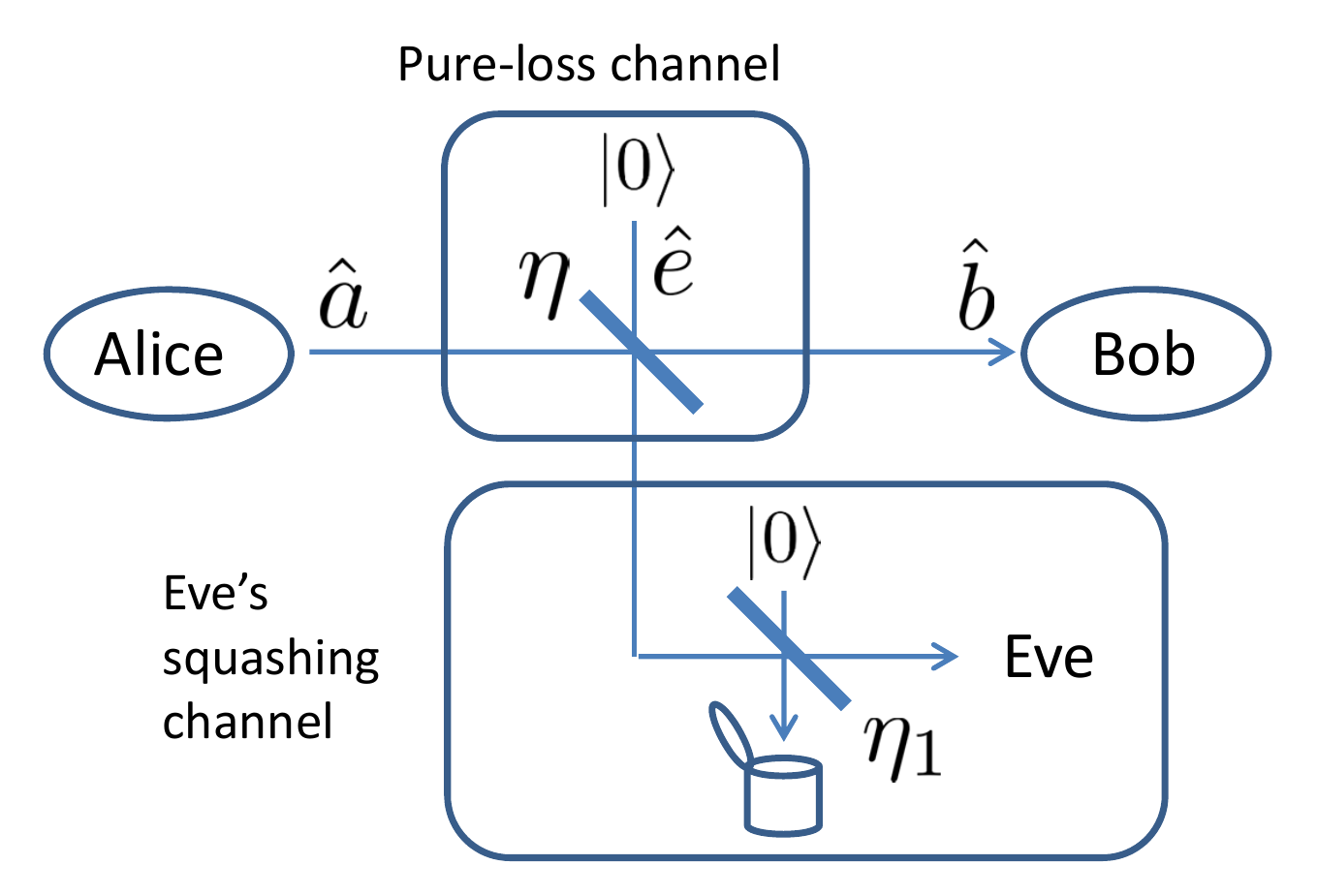}
\caption{{\bf Setup for calculating the upper bound on the secret key rate 
of the pure-loss optical channel}}. 
\label{fig:bosonic_setup}
\end{figure}

{\bf Pure-loss optical channel.}
Now we are in a position to derive a limit on the key generation rate of any QKD protocol which uses a lossy optical channel. The following input-output relation models linear loss in the propagation of an optical mode, such as through a lossy fiber or free space:%
\[
\hat{b}=\sqrt{\eta}\,\hat{a}+\sqrt{1-\eta}\,\hat{e},
\]
where $\hat{a}$, $\hat{b}$, and $\hat{e}$ are bosonic mode
operators corresponding to the sender Alice's input, the receiver Bob's output, and the environmental input, respectively. For the pure-loss bosonic channel, the environment mode is in its vacuum state. The transmittance of the channel, $\eta\in\left[  0,1\right]  $ is the fraction of input photons that makes it to the output on average. Let $\mathcal{N}_{\eta}$ denote the
channel from Alice to Bob. For a secret-key agreement protocol assisted by two-way classical-communication over this channel, we assume that it begins and ends with finite-dimensional states, but the processing between the first and final step can be conducted with infinite-dimensional systems 
(see the Methods section for further discussion of this point).
Furthermore, as is common in bosonic channel analyses \cite{GGLMSY04}, 
we begin by imposing a mean input power constraint. That is, for each input mode, we require that $\left\langle \hat{a}^{\dag}\hat{a}\right\rangle \leq N_{\rm S}$, with $0\leq N_{\rm S}<\infty$.
Thus, $E_{\text{sq}}\left(  \mathcal{N}_{\eta}\right)  $, with the additional photon number constraint on the channel input is an upper bound on $P_{2}\left(\mathcal{N}_{\eta}\right)$. 
By taking the squashing channel for the environment Eve to be
another pure-loss bosonic channel of transmittance $\eta_{1}\in\left[0,1\right]$ (see Fig.~\ref{fig:bosonic_setup}), noting that the resulting conditional mutual information can be written as a sum of two conditional entropies, and applying the extremality of Gaussian states with respect to conditional entropies~\cite{EW07,WGC06}, we find that the following quantity serves as an upper bound on $E_{\text{sq}}\left(  \mathcal{N}_{\eta
}\right)  $ for all $\eta_{1}\in\left[  0,1\right]  $ (see Supplementary Note 2 for a detailed proof):%
\begin{multline}
\tfrac{1}{2}\Big[g\left(  \left(  1-\eta_{1}+\eta\eta_{1}\right)
N_{\rm S}\right)  +g\left(  \left(  \eta_{1}+\eta\left(  1-\eta_{1}\right)
\right)  N_{\rm S}\right) \label{eq:bosonic-upper-bounds}\\
-g\left(  \eta_{1}\left(  1-\eta\right)  N_{\rm S}\right)  -g\left(  \left(
1-\eta_{1}\right)  \left(  1-\eta\right)  N_{\rm S}\right)  \Big],
\end{multline}
where $g\left(  x\right)  \equiv\left(  x+1\right)  \log_{2}\left(
x+1\right)  -x\log_{2}x$ is the Shannon entropy of a geometric distribution with mean $x$. The function $g(x)$ is also equal to the von Neumann entropy of a zero-mean circularly-symmetric thermal state with mean photon number $x$. Since the function in (\ref{eq:bosonic-upper-bounds}) is symmetric and convex in $\eta_{1}$, its minimum occurs at $\eta
_{1}=1/2$, leading to the following simpler upper bound:%
\[
g\left(  \left(  1+\eta\right)  N_{\rm S}/2\right)  -g\left(  \left(
1-\eta\right)  N_{\rm S}/2\right)  .
\]
By taking the limit of this upper bound as $N_{\rm S}\rightarrow\infty$, 
we obtain the photon-number independent expression,
\[
\log_{2}\left(  \left(  1+\eta\right)  /\left(  1-\eta\right)  \right)  ,
\]
which recovers the upper bound stated in \eqref{eq:good-upper-bound}. 

As mentioned in the introduction, 
any excess noise in the channel can only reduce 
the squashed entanglement of a quantum channel 
and thus \eqref{eq:good-upper-bound} serves as a fundamental upper limit 
on the secret key agreement capacity of a lossy optical channel. 
This statement follows from a quantum data processing argument,  
i.e., the quantum conditional mutual information does not increase under 
processing (including noise additions) of one of the systems that is not 
the conditioning system (see Proposition 3 of \cite{CW04}). 
Note that the statement does not prohibit the improvement 
of the key rates by applying `noisy processing' 
in specific existing QKD protocols such as BB84 as proposed 
in \cite{RGK05,RS07}. However, such an improved key rate is always 
equal or lower than our bound in \eqref{eq:good-upper-bound}. 

\begin{figure}[!t]
\centering
\includegraphics[width=3.0in]{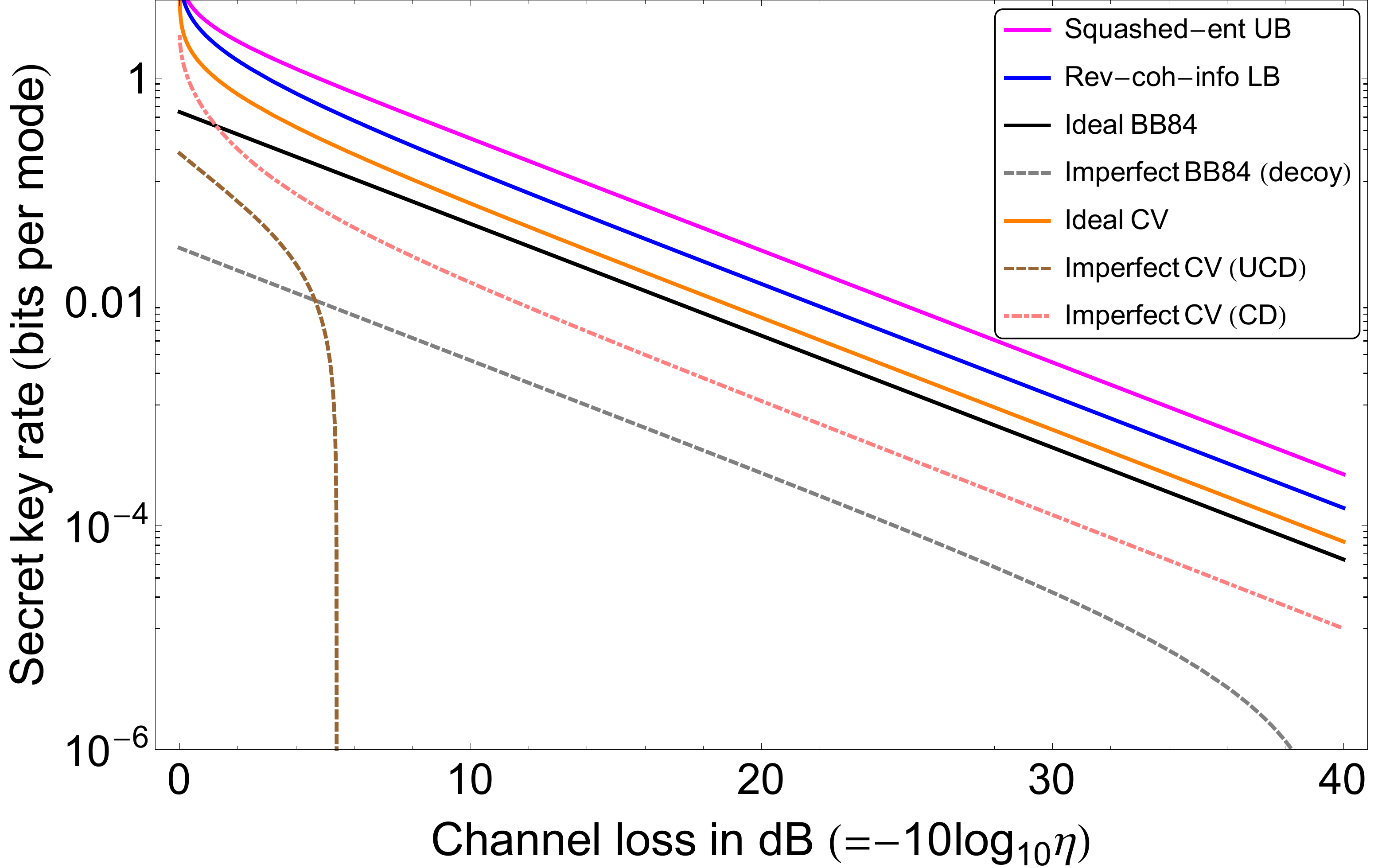}
\caption{
{\bf Upper and lower bounds on $P_2$ for a pure-loss bosonic channel}. 
The magenta solid curve is the squashed entanglement upper bound in 
(\ref{eq:good-upper-bound}). 
The blue solid curve is the reverse coherent information lower bound. 
The black solid curve is the efficient BB84 protocol with perfect devices 
and single photons, and the gray dotted curve is the decoy BB84 protocol 
including experimental imperfections \cite{SBCDLP09}. 
The orange solid curve is the Gaussian modulated coherent state 
continuous variable protocol (CV-GG02) \cite{GG02} with perfect devices. 
The brown dashed and pink dash-dotted curves are the CV-GG02 protocol 
including experimental imperfections with the uncalibrated- and 
calibrated-device scenarios, respectively. 
The details of the protocols and device parameters are described in 
the Methods section. 
}
\label{lossybosonic}
\end{figure} 

\section*{Discussion}
It is instructive to compare our upper bound to the known lower bounds on the secret-key agreement capacity of optical QKD protocols. 
BB84 is the most widely examined QKD protocol. 
When operating under polarization 
encoding and ideal conditions over a lossy channel 
(perfect single photon sources and detectors, 
and with the efficient BB84 protocol as proposed in~\cite{LCA06}), 
and with no excess noise (i.e., Eve can do only passive attacks 
consistent with the channel loss alone), the key rate 
is simply given by $\eta/2$ secret key bits per mode~\cite{SBCDLP09,LCA06}. 
The best known lower bound on the secret-key agreement capacity 
of the pure-loss channel ${\mathcal N}_\eta$ was established 
in Ref.~\cite{PGBL09}: 
\begin{equation}
P_2(\mathcal{N}_\eta) \ge \log_{2}\left(  1/\left(  1-\eta\right)  \right)\;{\text{key bits per mode}}.
\label{eq:lower-bound-p2}%
\end{equation}
These lower bounds and our upper bound are compared in Fig.~\ref{lossybosonic}, where 
we also plot the rate achievable with coherent-state continuous variable 
protocol (CV-GG02) \cite{SBCDLP09,GG02}, another major protocol, 
without any excess noise or imperfections. 
Also, as examples of the practical performance of QKD, 
we plot the decoy-state BB84 protocol including device imperfections 
as well as the imperfect CV-GG02 with uncalibrated- and 
calibrated-device scenarios 
(see the Methods section for the details of the protocols and parameters).

We note the following important observations. First, the two bounds in (\ref{eq:good-upper-bound}) and (\ref{eq:lower-bound-p2}) become close for $\eta\ll1$ (the high-loss regime, relevant for long-distance QKD). Thus, for small $\eta$, our upper bound demonstrates that the protocol from \cite{PGBL09} achieving the lower bound in (\ref{eq:lower-bound-p2}) is nearly optimal. To be precise, the upper and the lower bounds are well approximated by $2\eta/\ln 2$ and $\eta/\ln 2$ key bits per mode, when $\eta \ll 1$ (see Fig.~\ref{lossybosonic}). Furthermore, the ideal BB84 rate ($\eta$ key bits per mode) is worse than the reverse coherent information lower bound, only by a constant ($2/\ln 2 \approx 2.88$) factor in the high-loss regime where the factor 2 reflects the fact that the BB84 uses two polarization (or other) modes to send one bit and $\ln 2$ reflects some kind of gap between the qubit and continuous variable protocols. On the other hand, the protocol described in~\cite{PGBL09} that attains the reverse coherent information rate, requires an ideal SPDC source, and collective quantum operations and measurements, structured realizations of which are not known. 
In addition, even with detector impairments, both the decoy-state BB84 
as well as the CV-GG02 protocol (with or without the assumption 
of calibrated devices) can achieve secret key rates that scale linearly 
with the channel transmittance, until a maximum channel loss threshold 
where the rate plummets to zero. For BB84, this loss value where 
the rate-cliff occurs is driven by the detector dark counts, 
whereas for CV-GG02, it is driven primarily by the electronic noise 
in Bob's homodyne detector.
Hence, given the comparisons shown in Fig.~\ref{lossybosonic}, and since BB84 and CV protocols are realizable with currently available resources, it does not seem very worthwhile to pursue alternative repeater-less QKD protocols for 
higher key generation rate over a lossy channel.

Second, our bound significantly advances one of the long-standing open problems in quantum information theory, that of finding a good upper bound on $P_2(\mathcal{N})$, as well as for the two-way assisted quantum capacity $Q_2(\mathcal{N})$ (number of qubits that can be sent perfectly per use of a quantum channel with two-way classical-communication assistance). This is true
for a general quantum channel ${\mathcal N}$, 
and in particular for optical quantum channels such as 
$\mathcal{N}_\eta$. 
One of the important open questions is 
whether or not the true $P_2(\mathcal{N}_\eta)$ is additive. 
In other words, the question is whether the protocol that attains 
$P_2(\mathcal{N}_\eta)$ requires an input state that is entangled over 
several channel uses, or if a product input state suffices. 
In general, it is likely that $P_2(\mathcal{N})$ is super-additive 
for some channel as is the case for the unassisted secret-key agreement capacity 
$P(\mathcal{N})$~\cite{LWZG09} and the classical capacity 
of quantum channels~\cite{H09}. 
On the other hand, it is known that the classical capacity and
the unassisted quantum capacity of 
the lossy optical (bosonic) channel are additive~\cite{GGLMSY04,WPG07}. 
As mentioned above, our upper bound on $P_2(\mathcal{N})$ 
is a single-letter expression for any channel, i.e., the input optimization 
to evaluate our upper bound needs to be performed over a single channel use. 
The lower bound~\eqref{eq:lower-bound-p2} is the single-letter reverse 
coherent information evaluated for the lossy bosonic channel, 
whose operational interpretation is an entanglement distribution rate 
achievable via a product input realizable using a two-mode squeezed vacuum 
state, and collective quantum operations in the subsequent steps 
of the protocol, which uses classical feedback~\cite{GPLS09}. 
Thus, despite the fact that the additivity question for 
$P_2(\mathcal{N}_\eta)$ remains open, any super-additive gain cannot be 
very large in the high loss regime, and $P_2(\mathcal{N}_\eta)$ 
must scale as $\sim \eta$, when $\eta \ll 1$.

As a final point, consider a two-way QKD protocol, i.e., when Alice and Bob may use the lossy optical channel ${\mathcal N}_\eta$ in both directions, and also communicate freely over a two-way public channel. In such a case, the secret-key agreement capacity is upper bounded by $2\log_2\left((1+\eta)/(1-\eta)\right)$ secret key bits per mode transmitted in both directions.

In summary, we have established in \eqref{eq:good-upper-bound} an upper bound on the rate at which any QKD protocol can generate a shared secret key over a point-to-point lossy optical channel. This upper bound is a function of the channel loss only, and it does not increase with increasing transmit power. We compared our upper bound with the best known lower bound on the key rate and a key rate attainable in principle by the BB84 and CV-GG02 protocols under ideal operating conditions. This comparison reveals that there is essentially no scaling gap between the rates of known protocols and the ultimate secret-key agreement capacity, and thereby that there is no escaping the fundamental exponential decay of key rate with distance. The result of this paper on the one hand provides a powerful new tool for upper bounding the private capacity with two-way classical communication assistance, for a general quantum channel. 
On the other hand, the application to QKD over optical channels 
strongly suggests the need for quantum repeaters to perform QKD 
at high rates over long distances, no matter which actual QKD protocol 
one may choose to use.

Some important open questions remain. 
Although our bound applies for any finite number of channel uses,
one might be able to improve upon our result by establishing
a strong converse theorem or even better by considering a second-order 
analysis, along the lines discussed in \cite{TH12}. 
For establishing a strong converse theorem, some combination of the ideas 
presented in \cite{BBCW13,Oppenheim08} might be helpful. 
Another important open question is whether our upper bound
in (\ref{eq:good-upper-bound}) could be achievable using some QKD protocol, or whether the bound can be further tightened by choosing a squashing channel for Eve other than
the pure-loss channel (as shown in Fig.~\ref{fig:bosonic_setup}) 
or by investigating upper bounds other than $E_{\rm sq}(\mathcal{N})$. 
For this last question, some recent results in classical information theory 
\cite{GA10-1,GA10-2} might be helpful.

\section*{Methods}

{\bf Weak converse and the key rate upper bound for a finite number of channel uses.}
Although our main result establishes only a weak converse theorem, it is possible to estimate 
the effect of a finite number of channel uses, which is always the case in any practical 
setting. We carefully estimate $f(\varepsilon)$ 
discussed in the proof of Theorem 1. 
From the continuity inequality in (\ref{eq:continuity}), 
we can explicitly describe the additional term $f(\varepsilon)$: 
\begin{equation*}
nR \le n E_{\rm sq}(\mathcal{N}) + 16\sqrt{\varepsilon} \log d 
+ 4 h_2\left(2\sqrt{\varepsilon}\right) ,
\end{equation*}
where $d = 2^{nR}$. This implies the following bound:
\begin{equation*}
R \le \frac{1}{1-16\sqrt{\varepsilon}} \left(
E_{\rm sq}(\mathcal{N}) + 
4 h_2\left(2\sqrt{\varepsilon}\right)/n \right).
\end{equation*}
In the limit of large $n$, the second term
$4 h_2\left(2\sqrt{\varepsilon}\right)/n$ vanishes and the upper bound becomes
\begin{equation*}
R \le \frac{1}{1-16\sqrt{\varepsilon}} \left(
E_{\rm sq}(\mathcal{N})\right),
\end{equation*}
which suggests that there might be room for a trade-off between
communication rate and error probability/security
as quantified by $\varepsilon$.
If one could establish a strong converse theorem, this would eliminate
the implied trade-off given above, in the ideal case showing that the bound
$R \le  E_{\rm sq}(\mathcal{N})$ would hold in the large $n$ limit regardless
of the value of $\varepsilon$.

Let us consider a quantitative example, consisting of a pure-loss channel 
with $\varepsilon = 10^{-10}$ and $n=10^4$ (these are not too far 
from realistic QKD parameters \cite{TOKYO_QKD}). For these values, we get
\begin{align}
1 / (1-16\sqrt{\varepsilon}) & \approx 1.0002 , \\
4 h_2\left(2\sqrt{\varepsilon}\right)/n & \approx 1.36 \times 10^{-7}.
\end{align}
Furthermore, a 200km fiber with $0.2$dBkm$^{-1}$ 
corresponds to $\eta =10^{-4}$ 
and $\log ((1+\eta)/(1-\eta)) \approx 2.885 \times 10^{-4}$.
Replacing $E_{\rm sq}(\mathcal{N})$ with our upper bound 
$\log ((1+\eta)/(1-\eta))$ (see \eqref{eq:good-upper-bound}) 
and plugging in the above values of $\varepsilon$ and $n$, we find that
\begin{equation*}
R \leq 2.887 \times 10^{-4},
\end{equation*}
which is rather close to $\log ((1+\eta)/(1-\eta)) \approx 2.885 \times 10^{-4}$.
However, one can improve upon our upper bound by
establishing a strong converse theorem or even better by providing
a refined second-order analysis along the lines discussed in \cite{TH12}.

{\bf Infinite-dimensional system.} 
An optical channel can transmit infinite-dimensional 
(i.e., continuous variable) quantum states while Theorem 1 implicitly assumes 
finite-dimensional systems. 
We can circumvent this subtlety by assuming the protocol between Alice and 
Bob begins and ends with finite-dimensional states, but the processing between the first and final step can be conducted with infinite-dimensional systems. 
That is, their objective is to generate a maximally
entangled state $\left\vert \Phi\right\rangle _{AB}$\ or a finite number of
secret key bits, and they do so by Alice encoding a finite-dimensional quantum
state into an infinite-dimensional system and the final step of the protocol
has them truncate their systems to be of finite dimension. In this way, the
continuity inequality in the proof of Theorem 1 safely
applies and all of the other steps in between involve only the quantum data
processing inequality, which has been proven to hold in the general
infinite-dimensional setting \cite{U77}.

{\bf Decoy state BB84 and CV-GG02 protocols 
with experimental imperfections.}
The asymptotic secret key rates of the decoy state BB84 protocol and 
the CV-GG02 in Fig.~\ref{lossybosonic} are calculated 
from the theoretical models including imperfections 
summarized in \cite{SBCDLP09}. 
The parameters used for the plots are as follows: 
For the decoy BB84, the visibility of interference at Bob's receiver is 0.99, 
the transmittance of Bob's device is unity, the detector efficiency 
is 0.2, dark count rate is $10^{-6}$,  and 
the information leakage parameter due to the practical error code 
is set to be 1.2. 
For the CV-GG02 protocol, 
the optical noise is 0.005, the detector efficiency is 0.5, 
the electronic noise of the detector is 0.01, and the efficiency of 
the error correction code is set to be 0.9. 
These parameters are chosen to reflect the state of the art 
device technologies. 
In the `uncalibrated-device' scenario, Eve is able to access Bob's 
homodyne detector imperfections, e.g. to entangle the loss and noise 
of the detector. The `calibrated-device' scenario is based on the assumption 
that the homodyne detector is calibrated in a secure laboratory such that 
Eve cannot entangle her system to the detector imperfections. 
This assumption allows Alice and Bob to significantly extend the key rate 
and the loss tolerance. 
Note that the purpose of these plots is to compare these protocols under 
imperfections with our fundamental 
upper bound but is not to compare the practical aspects between these 
protocols (for example, our model does not include important practical 
conditions such as phase stability, repetition rate of the system, 
actual coding strategies, etc). 
For completeness, the key rate formulae for each protocol and scenario 
are described in Supplementary Note 3.

{\bf Acknowledgements}. We are grateful to Francesco Buscemi, Mikio Fujiwara, Min-Hsiu Hsieh, Seth Lloyd, Cosmo Lupo, Kiyoshi Tamaki, and Andreas Winter for insightful discussions. 
We also acknowledge Mark Byrd, Eric Chitambar, and the other participants of the Boris Musulin Workshop on Open Quantum Systems and Information Processing for helpful feedback. 
Finally, we thank Bob Tucci for kindly pointing us to his related work on squashed entanglement. 
This research was supported by the DARPA Quiness Program 
through US Army Research Office award W31P4Q-12-1-0019.
MT was partially supported by Open Partnership Joint Projects of 
JSPS Bilateral Joint Research Projects. 
SG acknowledges partial support from the SeaKey program, through 
the US Office of Naval Research contract number N00014-14-C-0002.

{\bf Author contributions}. All authors contributed to this work 
extensively and the write of the manuscript. 

{\bf Competing financial interests}. The authors declare no competing 
financial interests.

\clearpage

\begin{widetext}

\section{Supplementary note 1: 
Proof of the subadditivity inequality (Lemma 2)}
\label{SI_lemma2}
Let us restate the statement of the lemma:
For any five-party pure state $\psi_{AB_{1}E_{1}B_{2}E_{2}}$, the
following subadditivity inequality holds%
\begin{equation}
E_{\operatorname{sq}}\left(  A;B_{1}B_{2}\right)  _{\psi} \leq
E_{\operatorname{sq}}\left(  AB_{2}E_{2};B_{1}\right)  _{\psi} 
+ E_{\operatorname{sq}}\left(  AB_{1}E_{1};B_{2}\right)  _{\psi}.
\nonumber
\end{equation}

\begin{proof}
Let%
\begin{align*}
\tau_{AB_{1}E_{1}^{\prime}B_{2}E_{2}}  &  \equiv\mathcal{S}_{E_{1}\rightarrow
E_{1}^{\prime}}(\psi_{AB_{1}E_{1}B_{2}E_{2}}),\\
\sigma_{AB_{1}E_{1}B_{2}E_{2}^{\prime}}  &  \equiv\mathcal{S}_{E_{2}%
\rightarrow E_{2}^{\prime}}(\psi_{AB_{1}E_{1}B_{2}E_{2}}),\\
\omega_{AB_{1}E_{1}^{\prime}B_{2}E_{2}^{\prime}}  &  \equiv(\mathcal{S}%
_{E_{1}\rightarrow E_{1}^{\prime}}\otimes\mathcal{S}_{E_{2}\rightarrow
E_{2}^{\prime}})(\psi_{AB_{1}E_{1}B_{2}E_{2}}),
\end{align*}
where each $\mathcal{S}_{E_{i}\rightarrow E_{i}^{\prime}}$ is an arbitrary
local squashing channel. Let $\left\vert \phi^{\omega}\right\rangle
_{AB_{1}E_{1}^{\prime}B_{2}E_{2}^{\prime}R}$ be a purification of $\omega$
with purifying system $R$. The inequality in Lemma 2 is a
consequence of the following chain of inequalities:%
\begin{align*}
2E_{\text{sq}}\left(  A;B_{1}B_{2}\right)  _{\psi}  &  \leq I\left(
A;B_{1}B_{2}|E_{1}^{\prime}E_{2}^{\prime}\right)  _{\omega}\\
&  =H\left(  B_{1}B_{2}|E_{1}^{\prime}E_{2}^{\prime}\right)  _{\omega
}-H\left(  B_{1}B_{2}|E_{1}^{\prime}E_{2}^{\prime}A\right)  _{\omega}\\
&  =H\left(  B_{1}B_{2}|E_{1}^{\prime}E_{2}^{\prime}\right)  _{\phi}+H\left(
B_{1}B_{2}|R\right)  _{\phi}\\
&  \leq H\left(  B_{1}|E_{1}^{\prime}\right)  _{\phi}+H\left(  B_{2}%
|E_{2}^{\prime}\right)  _{\phi}\\
&  \ \ \ \ \ \ +H\left(  B_{1}|R\right)  _{\phi}+H\left(  B_{2}|R\right)
_{\phi} \\
&  =H\left(  B_{1}|E_{1}^{\prime}\right)  _{\omega}-H\left(  B_{1}|AB_{2}%
E_{1}^{\prime}E_{2}^{\prime}\right)  _{\omega}\\
&  \ \ \ \ \ \ +H\left(  B_{2}|E_{2}^{\prime}\right)  _{\omega}-H\left(
B_{2}|AB_{1}E_{1}^{\prime}E_{2}^{\prime}\right)  _{\omega}\\
&  =I\left(  AB_{2}E_{2}^{\prime};B_{1}|E_{1}^{\prime}\right)  _{\omega
}+I\left(  AB_{1}E_{1}^{\prime};B_{2}|E_{2}^{\prime}\right)  _{\omega}\\
&  \leq I\left(  AB_{2}E_{2};B_{1}|E_{1}^{\prime}\right)  _{\tau}+I\left(
AB_{1}E_{1};B_{2}|E_{2}^{\prime}\right)  _{\sigma}.
\end{align*}
The first inequality follows from the definition in (2). 
The first equality is a rewriting of the
conditional mutual information. The second equality exploits duality of
conditional entropy: for any pure tripartite state on systems $KLM$, the
equality $H\left(  K|L\right)  +H\left(  K|M\right)  =0$ holds. The second
inequality results from several applications of strong subadditivity (SSA) of
quantum entropy (SSA is the statement that $I\left(  K;L|M\right)  \geq0$ for
an arbitrary state on systems $KLM$) \cite{LR73}. The third equality again
exploits duality of conditional entropy and the last equality is just a
rewriting in terms of conditional mutual informations. The final inequality is
a result of a quantum data processing inequality for conditional mutual
information (see the proof of Proposition~3\ of \cite{CW04}). Since the
calculation above is independent of the choice of the maps $\mathcal{S}%
_{E_{i}\rightarrow E_{i}^{\prime}}$, the system $E_1$ purifies the state
on $A B_1 B_2 E_2$, and the system $E_2$ purifies the state on
$A B_1 B_2 E_1$, the subadditivity inequality in the
statement of the theorem follows.
\end{proof}

\section{Supplementary note 2: Squashed entanglement upper bound 
for the pure-loss bosonic channel}
\label{SI_bosonic_sq_channel}
Here we detail a proof that (5) is an upper bound
on $P_{2}\left(\mathcal{N}_{\eta}\right)$, 
where $\mathcal{N}_{\eta}$ is a pure-loss bosonic channel
with transmittance $\eta\in\left[  0,1\right]  $.

As mentioned before, we need to consider only pure states $\left\vert
\phi\right\rangle _{AA^{\prime}}$ when optimizing the squashed entanglement of
a quantum channel. Let $U_{E\rightarrow E^{\prime}F}^{\mathcal{S}}$ be an
isometric extension of Eve's squashing channel $\mathcal{S}_{E\rightarrow
E^{\prime}}$. Let $|\psi\rangle_{ABE^{\prime}F}\equiv U_{E\rightarrow
E^{\prime}F}^{\mathcal{S}}U^{\mathcal{N}}_{A^{\prime}\rightarrow BE}\left\vert
\phi\right\rangle _{AA^{\prime}}$, so that, $\mathrm{Tr}_{F}[|\phi
\rangle\langle\phi|_{ABE^{\prime}F}]=\mathcal{S}_{E\rightarrow E^{\prime}%
}\circ U_{A^{\prime}\rightarrow BE}^{\mathcal{N}}(\phi_{AA^{\prime}})$. Then%
\begin{align}
& \sup_{\phi_{AA^{\prime}}}E_{\text{sq}}(A;B)_{\mathcal{N}_{A^{\prime}\rightarrow
B}(\phi_{AA^{\prime}}%
)}\nonumber\label{eq:squashed_entanglement_pure_input}\\
& =\sup_{\phi_{AA^{\prime}}}\frac{1}{2}\inf_{\mathcal{S}_{E\rightarrow
E^{\prime}}}I(A;B|E^{\prime})\nonumber\\
& =\sup_{\phi_{AA^{\prime}}}\frac{1}{2}\inf_{\mathcal{S}_{E\rightarrow
E^{\prime}}}\left(  H(B|E^{\prime})_{\psi}-H(B|AE^{\prime})_{\psi}\right)
\nonumber\\
& =\sup_{\phi_{AA^{\prime}}}\frac{1}{2}\inf_{\mathcal{S}_{E\rightarrow
E^{\prime}}}\left(  H(B|E^{\prime})_{\psi}+H(B|F)_{\psi}\right)  .
\end{align}
The first equality follows from the definition of the squashed entanglement.
The second equality follows from the definition of conditional quantum mutual
information. The third equality uses the duality of conditional entropy and
the fact that $|\psi\rangle$ is pure.

Now suppose that Alice and Bob are connected by a pure-loss bosonic channel
with transmittance $\eta$. It is not necessarily an easy task to optimize
Eve's squashing channel $\mathcal{S}$. Instead, we consider a specific
squashing channel: a pure-loss bosonic channel $\mathcal{L}_{\eta_1}$ with
transmittance $\eta_{1}$. As shown above, the squashed entanglement can be
written as a sum of two conditional entropies, each of which is a function of
the reduced state Tr$_{A}\left\{  \phi_{AA^{\prime}}\right\}  $ on $A^{\prime}%
$. Since the overall channel from $A^{\prime}$ to $BE^{\prime}$ is Gaussian
and the overall channel from $A^{\prime}$ to $BF$ is Gaussian and due to
the photon-number constraint at the input, it follows from
the extremality of Gaussian states for conditional entropy \cite{EW07,WGC06}%
\ that a thermal state on $A^\prime$ of mean photon number $N_{\rm S}$ maximizes both of these
quantities.
With this and the fact that $\phi_{AA'}$ is a pure state,
we can conclude that the optimal $\phi_{AA'}$ is a two-mode
squeezed vacuum (TMSV) state. Let $N_{\rm S}$ be the average photon number of one share of the TMSV.
Then the covariance matrix of the reduced thermal state at $A'$ is given by
\[
\gamma^{A^{\prime}}=\left[
\begin{array}
[c]{cc}%
1+2N_{\rm S} & 0\\
0 & 1+2N_{\rm S}%
\end{array}
\right]  .
\]
Note that the covariance matrix is defined such that a vacuum state (or
coherent state) is described by an identity matrix. Therefore a covariance
matrix of the initial state in system $A^{\prime}E^{\prime}F$ is given by
$\gamma^{A^{\prime}}\oplus I^{E^{\prime}}\oplus I^{F}$. The beamsplitting
operations are given by the transformation
\[
\gamma^{A^{\prime}}\oplus I^{E^{\prime}}\oplus I^{F}\rightarrow S_{\eta_{1}%
}S_{\eta}\left(  \gamma^{A^{\prime}}\oplus I^{E^{\prime}}\oplus I^{F}\right)
S_{\eta}^{T}S_{\eta_{1}}^{T},
\]
where
\[
S_{\eta}=\left[
\begin{array}
[c]{ccc}%
\sqrt{\eta} & \sqrt{1-\eta} & 0\\
-\sqrt{1-\eta} & \sqrt{\eta} & 0\\
0 & 0 & 1
\end{array}
\right]  ^{\oplus2},\quad S_{\eta_{1}}=\left[
\begin{array}
[c]{ccc}%
1 & 0 & 0\\
0 & \sqrt{\eta_{1}} & \sqrt{1-\eta_{1}}\\
0 & -\sqrt{1-\eta_{1}} & \sqrt{\eta_{1}}%
\end{array}
\right]  ^{\oplus2},
\]
(the superscript \textquotedblleft$\oplus2$\textquotedblright\ means that the
same matrix is applied to both $x$ and $p$ quadratures. Because of the
symmetry of the state and the beamsplitter operation in phase space, basically
we need to consider only one quadrature.) This transformation is easily
calculated and we get a covariance matrix for the state $\mathrm{Tr}%
_{A}\left\{  |\phi\rangle\langle\phi|_{ABE^{\prime}F}\right\}  $:
\[
S_{\eta_{1}}S_{\eta}\left(  \gamma^{A^{\prime}}\oplus I^{E^{\prime}}\oplus
I^{F}\right)  S_{\eta_{1}}^{T}S_{\eta}^{T}=\left[
\begin{array}
[c]{ccc}%
1+\eta 2N_{\rm S} & -\sqrt{\eta(1-\eta)}\sqrt{\eta_{1}}2N_{\rm S} & \sqrt{\eta(1-\eta
)}\sqrt{1-\eta_{1}}2N_{\rm S}\\
-\sqrt{\eta(1-\eta)}\sqrt{\eta_{1}}2N_{\rm S} & 1+(1-\eta)\eta_{1}2N_{\rm S} &
-(1-\eta)\sqrt{\eta_{1}(1-\eta_{1})}2N_{\rm S}\\
\sqrt{\eta(1-\eta)}\sqrt{1-\eta_{1}}2N_{\rm S} & -(1-\eta)\sqrt{\eta_{1}(1-\eta
_{1})}2N_{\rm S} & 1+(1-\eta)(1-\eta_{1})2N_{\rm S}%
\end{array}
\right]  ^{\oplus2}.
\]
It immediately implies a covariance matrix of the marginal state
on $E^{\prime}$:
\[
\gamma_{E^{\prime}}=\left[
\begin{array}
[c]{cc}%
1+(1-\eta)\eta_{1}2N_{\rm S} & 0\\
0 & 1+(1-\eta)\eta_{1}2N_{\rm S}%
\end{array}
\right]  ,
\]
which is the covariance matrix for a thermal state with photon number $(1-\eta)\eta_{1}N_{\rm S}$. Thus we
have
\[
H(E^{\prime})=g\left(  (1-\eta)\eta_{1}N_{\rm S}\right)  ,
\]
where $g(x)=(1+x)\log(1+x)-x\log x$. Similarly, we get
\[
H(F)=g\left(  (1-\eta)(1-\eta_{1})N_{\rm S}\right)  .
\]
The other entropies $H(BE^{\prime})$ and $H(BF)$ are also obtained by
considering the corresponding submatrices and diagonalizing them. Then we can
find
\begin{align}
H(BE^{\prime})  & =g\left(  \{\eta+(1-\eta)\eta_{1}\}N_{\rm S}\right)  \nonumber\\
H(BF)  & =g\left(  \{\eta+(1-\eta)(1-\eta_{1})\}N_{\rm S}\right)
\end{align}
As a consequence, we obtain the upper bound,%
\begin{align}
P_{2}(\mathcal{N}_{\eta})  & \leq\min_{\eta_{1}}\frac{1}{2}\big\{g\left(
\{\eta+(1-\eta)\eta_{1}\}N_{\rm S}\right)  -g\left(  (1-\eta)\eta_{1}N_{\rm S}\right)
\nonumber\label{eq:lossy_bosonic_upper_bound1} +g\left(  \{\eta+(1-\eta)(1-\eta_{1})\}N_{\rm S}\right)  -g\left(
(1-\eta)(1-\eta_{1})N_{\rm S}\right)  \big\}\nonumber\\
& =g\left(  (1+\eta)N_{\rm S}/2\right)  -g\left(  (1-\eta)N_{\rm S}/2\right)  .
\end{align}
The minimal value is achieved by $\eta_{1}=1/2$ because the function is symmetric and convex in $\eta_1$ (with convexity checked by computing the second derivative).
The expression $g\left(  (1+\eta)N_{\rm S}/2\right)  -g\left(  (1-\eta)N_{\rm S}/2\right)  $ converges to
$\log(1+\eta)/(1-\eta)$ as $N_{\rm S}\rightarrow\infty$.

\section{Supplementary note 3: Secret key rates for the decoy BB84 
and CV-GG02 protocols}

In this supplementary note, we describe the asymptotic key rate 
formulae for the decoy BB84 and Gaussian modulated coherent-state 
CV-GG02 protocols. Detailed description of the protocols and 
the derivation of their key rates are for example found in \cite{SBCDLP09} 
and references therein. 
Note that we only consider the asymptotic limit, that is, 
the channel estimation is perfectly done by negligibly small 
amount of pulses.

{\bf Decoy BB84 protocol}.  
Let $\xi$ be the mean photon number of the pulse which is tunable to 
generate decoy states. For each $\xi$, the key rate is given by 
\begin{equation}
\label{eq:decoy_key_rate}
K^\xi = R^\xi \left\{ Y_0^\xi + Y_1^\xi ( 1-h(\epsilon_1) 
- {\rm leak}_{\rm EC}(Q^\xi) \right\} .
\end{equation}
We can assume that a given value $\mu$ is almost always used for $\xi$. 
Thus in the following we only consider $\xi=\mu$ and then optimize 
$\mu$ to maximize the key rate in Eq.~(\ref{eq:decoy_key_rate}). 
The parameters in the rhs of Eq.~(\ref{eq:decoy_key_rate}) are 
as follows: $R$ is the total detection rate (here it is given by 
the rate per mode per pulse. Note that the deocy BB84 usually uses two 
modes, such as polarization or time bins, to send one bit 
of quantum information), $Y_n=R_n/R$ where 
$R_n$ is the detection rate for the events when Alice sent $n$ photons 
($\sum_n R_n=R$ and thus $\sum_n Y_n =1$), 
$\epsilon_n$ is the error rate on the $n$ photon signal, 
$h(\cdot)$ is the binary entropy function, 
$Q$ is the quantum bit error rate (QBER), and 
${\rm leak}_{\rm EC}(Q^\xi) = f h(Q)$ and $f\ge 1$ represents the efficiency 
of the error correction code. 
For the decoy BB84 protocol, these terms are given by 
\begin{eqnarray}
R^\mu & = & \left(1-e^{-\mu t} (1-2p_{\rm d}) \right) (1-p_{\rm d}) /2 ,
\\
Y_0^\mu & = & \frac{2 p_{\rm d} e^{-\mu}}{1-e^{-\mu t} 
+ 2p_{\rm d} e^{-\mu t}} ,
\\
Y_1^\mu & = & \frac{\mu e^{-\mu} \left( t(1-p_{\rm d}) + 2 (1-t) p_{\rm d} 
\right)}{ 1-(1-2p_{\rm d}) e^{-\mu t} } ,
\\
\epsilon_1 & = & \frac{t \epsilon + (1-t) p_{\rm d} }{
t + 2 (1-t) p_{\rm d}}
\\
Q^\mu & = & \frac{\epsilon - (\epsilon-p_{\rm d}) e^{-\mu t}}{
1- (1-2p_{\rm d}) e^{-\mu t}}
\end{eqnarray}
where $t = \eta \eta_B \eta_{\rm d}$, $\eta$ is the channel transmittance, 
$\eta_B$ is the efficiency of Bob's devices, $\eta_{\rm d}$ is 
the efficiency of Bob's detectors. 
$p_{\rm d}$ is the dark counts of the detector, and 
$\epsilon = (1-V)/2$ with the visibility $V$. 
The optimal key rate is then given by 
\begin{equation}
K_{\rm opt}^{\rm decoy} = \max_\mu K^\mu .
\end{equation}

{\bf CV-GG02 protocol}. 
The key rate is given by 
\begin{equation}
K = R [\beta I(A;B) - \chi(B;E)] ,
\end{equation}
where $R$ is the repetition rate, which is omitted in our calculation, 
$\beta$ is the reconciliation efficiency, $I(A;B)$ is the mutual information 
between Alice and Bob, and $\chi(B;E)$ is the Holevo information between 
Bob and Eve. 
In the following we consider the uncalibrated- and calibrated-device 
scenarios. 
For both scenarios, $I(A;B)$ has the same expression:
\begin{equation}
I(A;B) = \frac{1}{2} \log_2 \frac{\delta+v}{\delta+1} ,
\end{equation}
where $v = v_A + 1$ is the variance of the Alice's output thermal state and 
$v_A$ is the variance of the Alice's Gaussian modulation. 
$\delta$ is the total noise between Alice and Bob: 
\begin{equation}
\delta = \delta_{\rm ch} + \delta_{\rm h}/\eta, 
\end{equation}
where $\delta_{\rm ch} = (1-\eta)/\eta + \epsilon$ is the channel noise 
with the channel tarnsmittance $\eta$ and the optical excess noise $\epsilon$, 
and $\delta_{\rm h} = (1+v_{\rm el})/\eta_{\rm d} -1$ 
is the noise of the Bob's homodyne detector with the homodyne detector's electronic noise $v_{\rm el}$ and quantum efficiency $\eta_{\rm d}$.

The amount of the Holevo information depends on the scenario. 
For the uncalibrated-device scenario, Eve can hold a whole environmental 
part including the loss and noise of the homodyne detector at Bob's side. 
Then the Holevo information is calculated to be \cite{SBCDLP09} 
\begin{equation}
\chi_{\rm uc}(B;E)=g(\tilde{\lambda}_1) + g(\tilde{\lambda}_2) 
- g(\tilde{\lambda}_3), 
\end{equation}
where $\tilde{\lambda}_k = (\lambda_k-1)/2$, 
\begin{eqnarray}
\lambda^2_{1,2} & = & \frac{1}{2}\left( A \pm \sqrt{A^2-4B} \right) ,
\\
\lambda^2_3 & = & \frac{v(1+v \delta)}{v+\delta} ,
\end{eqnarray}
and 
\begin{eqnarray}
A & = & v^2 (1-2\eta \eta_{\rm d}) + 2\eta \eta_{\rm d} 
+ \left\{ \eta \eta_{\rm d} (v+\delta) \right\}^2 ,
\\
B & = & \left\{ \eta \eta_{\rm d} (v \delta + 1) \right\}^2. 
\end{eqnarray}
The achievable key rate is then obtained by maximizing $K$ over $v \ge 1$. 

For the calibrated-device scenario, Eve cannot access to the homodyne 
detector's imperfections. The Holevo information for such a scenario 
is for example given in \cite{LBGFKDDCBMG07}. We summarize it here 
for the completeness: 
\begin{equation}
\chi_{\rm c}(B;E)=g(\tilde{\lambda}'_1) + g(\tilde{\lambda}'_2) 
- g(\tilde{\lambda}'_3) - g(\tilde{\lambda}'_4), 
\end{equation}
where $\tilde{\lambda}'_k = (\lambda'_k-1)/2$,  
\begin{eqnarray}
\lambda'^2_{1,2} & = & \frac{1}{2}\left( A' \pm \sqrt{A'^2-4B'} \right) ,
\\
\lambda'^2_{3,4} & = & \frac{1}{2}\left( C' \pm \sqrt{C'^2-4D'} \right) ,
\end{eqnarray} 
and
\begin{eqnarray}
A' & = & v^2 (1-2\eta) + 2\eta + \left\{ \eta(v+\delta_{\rm ch}) \right\}^2 ,
\\
B' & = & \left\{ \eta(v\delta_{ch}+1) \right\}^2 ,
\\
C' & = & \frac{v\sqrt{B'} + \eta (v+\delta_{\rm ch}) + A' \delta_{\rm h}}{
\eta (v+\delta)} ,
\\
D' & = & \frac{v\sqrt{B'} + B' \delta_{\rm h}}{\eta(v+\delta)} ,
\end{eqnarray}
The achievable key rate is again obtained by maximizing $K$ over $v \ge 1$.

\end{widetext}
\end{document}